# Momentum-resolved electronic structure at a buried interface from soft x-ray standing-wave angle-resolved photoemission

A. X. Gray,[1,2,3] J. Minár,[4] L. Plucinski,[5] M. Huijben,[6] A. Bostwick,[7] E. Rotenberg,[7] S.-H. Yang,[8] J. Braun,[4] A. Winkelmann,[9] G. Conti,[1,2] D. Eiteneer,[1,2] A. Rattanachata,[1,2] A. A. Greer,[1,2] J. Ciston,[10] C. Ophus,[10] G. Rijnders,[6] D. H. A. Blank,[6] D. Doennig,[11] R. Pentcheva,[11] C. M. Schneider,[5] H. Ebert,[4] and C. S. Fadley[1,2]

[1]*Department of Physics, University of California Davis, Davis, California 95616, USA*

[2]*Materials Sciences Division, Lawrence Berkeley National Laboratory, Berkeley, California 94720, USA*

[3]*Stanford Institute for Materials and Energy Science, Stanford University and SLAC National Accelerator Laboratory, 2575 Sand Hill Road, Menlo Park, California 94029, USA*

[4]*Department of Chemistry, Physical Chemistry Institute, Ludwig Maximillian University, München, Germany*

[5]*Peter-Grünberg-Institut PGI-6, Forschungszentrum Jülich GmbH, 52425 Jülich, Germany*

[6]*Faculty of Science and Technology, MESA$^+$ Institute for Nanotechnology, University of Twente, Enschede, The Netherlands*

[7]*Advanced Light Source, Lawrence Berkeley National Laboratory, Berkeley, California 94720, USA*

[8]*IBM Almaden Research Center, San Jose, California 95120, USA*

[9]*Max-Planck-Institut für Mikrostrukturphysik, Weinberg 2, D-06120 Halle (Saale), Germany*

[10]*National Center for Electron Microscopy, Lawrence Berkeley National Laboratory, Berkeley, California 94720, USA*

[11]*Department of Earth and Environmental Sciences and Center of Nanoscience (CENS), Ludwig Maximillian University, München, Germany*

PAC Numbers: 79.60.Jv, 68.49.Uv, 85.75.Dd

**Abstract** – Angle-resolved photoemission spectroscopy (ARPES) is a powerful technique for the study of electronic structure, but it lacks a direct ability to study buried interfaces between two materials. We address this limitation by combining ARPES with soft x-ray standing-wave (SW) excitation (SWARPES), in which the SW profile is scanned through the depth of the sample. We have studied the buried interface in a prototypical magnetic tunnel junction $La_{0.7}Sr_{0.3}MnO_3/SrTiO_3$. Depth- and momentum-resolved maps of Mn $3d$ $e_g$ and $t_{2g}$ states from the central, bulk-like and interface-like regions of $La_{0.7}Sr_{0.3}MnO_3$ exhibit distinctly different behavior consistent with a change in the Mn bonding at the interface. We compare the experimental results to state-of-the-art density-functional and one-step photoemission theory, with encouraging agreement that suggests wide future applications of this technique.

**Introduction--** Angle-resolved photoemission (ARPES) is the technique of choice for probing the electronic structure of solids and surfaces, yielding as direct output a map of photoelectron intensities as a function of electron kinetic energy $E_{kin}$ and electron momentum $\vec{p} = \hbar\vec{k}$, and it has been applied to virtually every type of crystalline material [1,2]. A typical experimental setup involving a hemispherical electrostatic analyzer is shown in fig. 1(a). [3]. For excitation with a photon energy $h\nu$, three-dimensional datasets of kinetic energy $E_{kin}(\vec{k})$ or binding energy relative to the Fermi level $E_b^F(\vec{k}) \approx h\nu - E_{kin}(\vec{k})$ as a function of the $k_x$ and $k_y$ components are obtained by measuring detector images of $E_{kin}$ versus the take-off angle $\theta_{TOA}$ and scanning also the orthogonal angle $\beta_{TOA}$ by rotating the sample. Each point in this volume can in turn be mapped into the reduced Brillouin zone (BZ) via direct transitions (DTs) that in their simplest form obey the conservation law $\vec{k} = \vec{k}_i + \vec{g}_{hk\ell} + \vec{k}_{h\nu}$, where $\vec{k}_i$ is in the reduced BZ, $\vec{g}_{hk\ell}$ is a bulk reciprocal lattice vector, and $\vec{k}_{h\nu}$ is the photon wave vector, which must be considered for energies in the soft and hard x-ray regimes above about 0.5 keV due to non-dipole effects [4,5].

However, a significant disadvantage of the conventional ARPES technique is its extreme surface-sensitivity, due to the very low inelastic mean-free paths (IMFPs) of the electrons photoemitted using radiation in the range 25 eV < $h\nu$ < 150 eV [6]. As a quantitative example, the IMFP, which is in turn the average normal emission depth, for the complex oxides $SrTiO_3$ or $La_{0.7}Sr_{0.3}MnO_3$ of interest here can be estimated to be about 1.9 Å at $h\nu$ = 25 eV, and 5.9 Å at $h\nu$ = 150 eV [6,7], or only a few atomic layers below the surface. A recent ARPES study of $La_{0.7}Sr_{0.3}MnO_3$ illustrates this surface sensitivity [8]. This has led in recent years to more bulk-sensitive ARPES measurements at higher photon energies and thus larger IMFPs in the 10-100 Å range that are by now being carried out in the soft x-ray regime of 500-1200 eV for various materials [4,5,9], as well as in the hard x-ray regime from 3.2 to 5.9 keV [12,13].

Yet, even at these higher photon energies, the photoemission signal originating closer to the surface will be stronger than the signal originating from below according to $I(z) = I_0 \exp[-z/\Lambda \sin\theta_{TOA}]$, where $z$ is the depth, $\Lambda$ is the IMFP, or more correctly the effective attenuation length (EAL) that includes elastic scattering effects as well [6] and $\theta_{TOA}$ is the electron takeoff angle relative to the surface (cf. fig. 1 (a) [6,7]). Controllable depth-selectivity can however be accomplished by setting-up an x-ray standing-wave (SW) field in the sample by growing it as, or on, a synthetic periodic multilayer mirror substrate, which in first-order Bragg reflection acts as a strong standing-wave (SW) generator [14,15]. The maxima of the SW can be moved in the $z$ direction perpendicular to the sample surface by scanning the incidence angle $\theta_{inc}$ through the Bragg condition, thus generating a well-known rocking curve (RC) of intensity [14,15,16]. Angle-integrated SW excited x-ray photoemission (SW-XPS) from core levels and valence bands has been applied previously in studies of various systems, in particular layers and interfaces of relevance to giant magnetoresistance (GMR) and tunnel magnetoresistance (TMR) [16], including the insulator/half-metallic ferromagnet system $SrTiO_3/La_{0.7}Sr_{0.3}MnO_3$ (STO/LSMO) that is the topic of this study [17], but $\vec{k}$-resolved ARPES has not previously been attempted. These prior SW-XPS studies were carried out at room temperature, such that phonon-induced non-direct transitions (NDTs) led to an averaging over the BZ, and resultant valence spectra that closely resemble matrix-element weighted densities of states (MEW-DOS) [4,5]. It is possible to estimate the fraction of direct transitions from a photoemission Debye-Waller factor of the form $W(T) = exp[-\frac{1}{3} g_{hk\ell}^2 \langle U^2(T) \rangle ]$, where $U^2(T)$ is the three-dimensional mean-squared vibrational displacement [4,5], and we consider this aspect further below.

In this letter, we add depth selectivity to ARPES by combining more bulk sensitive soft x-ray excitation at *ca.* 800 eV corresponding to IMFPs of about 19 Å with the SW approach (SWARPES) to provide a unique depth- and $\vec{k}$-resolved probe of buried layer and interface electronic structure. Interface electronic structure is known to be crucial to the properties of various nanoscale multilayer systems, as for example, in the magnetic tunnel junction (MTJ) Fe/MgO, for which the $\Delta_1$ band of Fe is thought to be the predominant carrier of spin-polarized tunneling current [18,19], and the interface between STO and LAO in the system LaAlO$_3$/SrTiO$_3$, for which the interface provides a 2D electron gas that has been shown to be both ferromagnetic and superconducting [20]. Yet there are up to now no techniques for directly studying <u>interface</u> electronic structure in a $\vec{k}$-resolved manner. We illustrate the capability of SWARPES to do this for a prototypical oxide MTJ, La$_{0.7}$Sr$_{0.3}$MnO$_3$/SrTiO$_3$ (LSMO/STO) by comparing experiment to theory of several types, including, in particular, state-of-the-art one-step photoemission calculations.

The much-studied LSMO/STO system is a promising candidate for a magnetic tunnel junction [21-23], wherein the half-metallic nature of ferromagnetic LSMO is responsible for producing a 100% spin-polarized tunneling current across the STO insulating barrier [24,25]. Up to now, however, the theoretically-predicted tunneling magnetoresistance (TMR) effect of 100% [26] has not been realized, with the highest TMR values reported so far being on the order of 80% [27-29]. The most widely accepted explanation for this reduced performance is highly-localized interface effects in the LSMO layer near the interface with STO [30]. In a prior angle-integrated SW-XPS study of LSMO/STO [17], we have investigated the chemical and electronic structure profiles of the LSMO/STO interface via *core-level* soft and hard x-ray standing-wave excited photoemission, x-ray absorption and x-ray reflectivity, in conjunction with x-ray optical [31] and core-hole multiplet theoretical modeling [32,33]. Analysis of the core-level standing-wave modulations revealed the presence of an interdiffusion region of 4-5 in thickness Å (a little over 1 unit cell) between the STO and LSMO layers, a change in the soft x-ray optical coefficients of LSMO near the interface, and a shift in the position of the Mn 3*p* peak near the interface that is consistent with a crystal-field distortion effect. What is still needed however is depth- <u>and</u> $\vec{k}$-resolved information concerning the valence electronic states. We show here that SWARPES can provide this.

The LSMO/STO multilayer sample consisted of 120 bilayers, each consisting of 4 unit cells of LSMO (~15.51 Å) and 4 unit cells of STO (~15.61 Å), with STO terminating the structure, as shown schematically in fig. 1(b), and was fabricated using the pulsed laser deposition (PLD) technique (see details in the Supplementary Information (SI) [34]). The transport and magnetic properties of the LSMO/STO superlattice are consistent with previous reports [35], as shown in our SI [34].

The SWARPES measurements were carried out at the Electronic Structure Factory (ESF) endstation at Beamline 7.0.1 of the Advanced Light Source (Lawrence Berkeley National Laboratory) using a Scienta R4000 spectrometer. The measurements were performed at a temperature of 20 K and with an overall energy resolution of ~300 meV. In order to maximize reflectivity and thus also the contrast of the standing-wave, and therefore to better define the depth-resolved photoemission within the sample, the excitation energy was set to 833.2 eV, which is just below the La $3d_{5/2}$ absorption edge, as discussed elsewhere [17].

In order to verify the presence of the SW in the superlattice, and to most quantitatively model the intensity profile of it within the sample, we first performed core-level SW-XPS measurements. Strong standing-wave RC intensity modulations near the Bragg condition for the superlattice were observed for Ti $2p_{3/2}$ and Mn $3p$ core-levels (solid curves in fig. 1(c); these are fully consistent with our prior study of a similar LSMO/STO sample [17]. These RCs were fitted using a specially-written theoretical code [31] in order to confirm the chemical profile of the structure with Ångstrom-level accuracy, and the best-fit theoretical curves are shown as dashed curves in fig. 1(c). The same x-ray optical model and sample configuration was used to simulate the electric-field intensity ($E^2$) profile of the standing wave inside the superlattice as a function of depth and incidence angle. The results of these simulations in fig. 1(d) reveal that the standing-wave maxima will highlight the center ("bulk") region of the buried LSMO layer at an incidence angle of 12.4° (a maximum of Mn $3p$ intensity), as shown in the left line-cut. Increasing the incidence angle past the Bragg condition, we shift the standing wave downwards by about half-a-period, highlighting the interfacial region of the LSMO layer at angles above 12.9° (a maximum of Ti $2p$ intensity), as shown in the right line-cut.

Three-dimensional SWARPES measurements were thus performed at these two incidence angles, as well as others, finally yielding $E_B^F(k_x, k_y(k_z))$, with $k_z$ implicitly known but not directly measured. To validate our final conclusions, we have also measured SWARPES at an additional five angles to the left (11.82°, 12.10°) and right (13.35°, 13.90°) of the rocking curve, as well as in the middle of it (12.15°); some of these results are presented in our SI [34].

In fig. 2(a) we show a typical $(k_x, k_y)$ photoemission intensity distribution obtained in the LSMO-bulk-sensitive geometry ($\theta_{inc} = 12.4°$) for a fixed binding energy of about -2.5 eV, which corresponds to the binding energy of Mn 3$d$ $t_{2g}$–derived states [25,36]. At this binding energy, we need not consider any contribution of the STO overlayer to spectra, since STO has the bandgap of ~3.25 eV, and therefore does not have states in this energy range [ref. 36, also see theoretical calculations in our SI (ref. 34)]. We do, however, have to consider that, to whatever degree this spectrum involves a mixture of $\vec{k}$-conserving direct transitions (DTs) and phonon-induced non-direct transitions (NDTs), the NDT components will have energy distributions reflecting the DOS and angular distributions corresponding to core-like x-ray photoelectron diffraction (XPD), as discussed elsewhere [4,5]. Correction for DOS and XPD effects can to first order be done by dividing the data successively by the average over angle and the average over energy of each detector image, respectively [37] as demonstrated for experimental data from W recently [12], and this has been done in arriving at the results shown in fig. 2(a). At 20 K, the fraction of DTs is estimated from the Debye-Waller factor to be ~75%, and thus NDTs, or in turn, XPD- and DOS-like effects, to be ~25% of the total intensity. In order to further separate the effects of true electronic-state dispersions from XPD, we have also measured the angle-resolved spectra of the Mn 3$p$ core-level concurrently with the valence-band measurements for each incidence angle. Core levels represent localized states with no dispersion in $\vec{k}$. Thus, the Mn 3$p$ pattern we see in fig. 2(b) is exclusively due to XPD, and this is confirmed by a dynamical Kikuchi-band XPD calculation shown in fig. 2(c) [38], which exhibits excellent agreement with the data of fig. 2(b). By now correcting the SWARPES spectrum in fig.

2(a) using the XPD-only spectrum in fig. 2(b), through a scaled division described in our SI [34], we can finally unambiguously isolate the $\vec{k}$ -resolved electronic structure of Mn 3$d$ $t_{2g}$-derived states. But we stress that our conclusions regarding interface-specific electronic state dispersions do not depend critically on this correction procedure, as discussed further in connection with difference data in our SI [34].

In fig. 3, we show some key SWARPES results for five key binding-energy positions, all corrected in the same two-step way to remove any DOS or XPD effects. The top panel (3(a)) shows a reference, angle-integrated spectrum spanning the Fermi-referenced binding-energy window from +1 eV to -9 eV, including five major features labeled 1 – 5; this curve should roughly represent the MEW-DOS for the sample. The region between the binding energies of 0 eV and -3.25 eV contains the LSMO-derived states, specifically Mn 3$d$ $e_g$ (feature 1, at ca. 1.0 eV) and Mn 3$d$ $t_{2g}$ (feature 2, at ca. 2.4 eV), and no STO-derived states due to the bandgap "window" [36, plus our SI (ref. 34)]. Conversely, the region between -3.25 eV and -7.0 eV we expect to be dominated by the states originating in the topmost STO layer (labeled 3,4, and expected to be relatively flat complex bands). The deeper bands from LSMO will also show up in this region, but with attenuated intensity due to the STO layer. However, we finally suggest that feature 5 is predominantly LSMO-derived, and show evidence below and in our SI [34] for this interpretation. Figs. 3(b), (c) and (d) now show the corrected low-temperature ARPES intensity maps in ($k_x$, $k_y$) summed over 300 meV intervals centered at binding energies 1-5 for (b) - the bulk-LSMO sensitive incidence angle, (c) - the interface-sensitive angle, and (d) - the bulk-minus-interface difference between the two. The contrast of the color map in the difference map of fig. 3(d) is enhanced in order to accentuate the smaller differences between the bulk-like and interface-like features. Note the indication of the surface normal and the first Brillouin zone in the image in fig. 3(b). Multiple Brillouin zones are thus represented in these images due to the large $k$ vector of a photoelectron at 833.2 eV excitation energy (14.8 Å$^{-1}$, compared to the BZ dimensions in STO or LSMO of 2π/a ≈1.61 Å$^{-1}$), and the fact that the detector images span 35° in $k_x$ and 40°in $k_y$.

The bulk- and interface- sensitive maps at a given energy are at first sight very similar, although those for different energies clearly differ markedly from one another. The most dispersive features 1, 2, and 5 show the most structure, the the less dispersive STO bands 3, and 4 much less structure. Feature 5 is striking in showing a remarkably simple square pattern that essentially represents the expected BZ repeat pattern; this simple appearance additionally confirms the validity of our DOS and XPD correction procedures. The bulk-minus-interface maps are finally crucial indicators of changes in the electronic structure at the interface. The biggest changes (up to ~4.5% intensity) are observed for the LSMO-derived Mn $3d$ $e_g$ and $t_{2g}$ electronic states, suggesting significant changes in the $(k_x, k_y)$ dependence of these states at the LSMO/STO interface and a general suppression of intensity. The STO-dominated states at the binding energies of 4.0 and 6.2 eV show less momentum dispersion in general, and thus also exhibit only minor bulk-interface changes. Finally, theory shown in our SI [34] indicates that the largely LSMO-derived states 5 at ~7.5 eV also exhibit a marked change in $(k_x, k_y)$ dependence near the interface. All of these changes, although subtle, represent a unique experimental insight into interface electronic structure, and we have verified the validity of these bulk-surface difference effects by making similar difference maps for points on either side of the rocking curve, which are found to show no discernible effects (see our SI [34]).

In comparing our experimental results to theory, we will consider only *k*-space maps for the LSMO-derived features 1 and 2 representing Mn $3d$ $t_{2g}$ and $e_g$ states, and feature 5 at the bottom of the valence bands that exhibits a very simple dispersion pattern [36]. Fig. 4(a) first presents the results of simple free-electron final-state calculations involving direct transitions from LSMO band-structure calculations performed using the Wien2k code at the local density approximation + U (LDA+U) level to allow for correlation effects [39]. Here the LSMO is assumed to be ferromagnetic and the spin of the photoelectrons is distinguished by color (red = majority, yellow = minority); thus the $e_g$ allowed transitions are all majority or red. The agreement between the experiment and theory is very encouraging for all three energies, with the Brillouin-zone periodicity and positions of some of the major features reproduced well, although of course there is no information in these *k*-space maps as to relative photoemission intensities, since no allowance is made for matrix elements. Analyzing feature 5 with the LSMO band structure is also found to be valid from LDA calculations for the full multilayer, which predict that VB minimum of LSMO to be below that of STO (see our SI [34]).

In figs. 4 (b)-(d), we present the results of much more accurate one-step photoemission theory based on a fully relativistic LDA+U layer-KKR (Korringa-Kohn-Rostoker) approach and a time-reversed LEED (low-energy electron diffraction) final-state [40] (further details in our SI [34]), as applied to the actual multilayer structure with the surface present. These calculations furthermore incorporate a first attempt to include the intensity profile of the x-ray standing wave by using as an additional input the $|E^2|$ profiles shown in fig. 1(d). The resulting bulk- and interface- sensitive photoemission intensity *k*-space maps are shown in figs. 4(b)-(c). Finally, differences between the bulk and interface electronic structure were calculated and these are plotted in fig. 4(d). It is important to note that this type of one-step theory calculation represents a much more accurate theoretical counterpart to this particular experiment as compared to free-electron final-state theory, since in addition to calculating true angular distributions of photoemission intensities due to SW excitation, the influence of the top STO overlayer is also taken into account, although in a rigid lattice approximation so that phonon effects are not included. Thus, although the periodicity of the Brillouin zones and the positions of the major features are similar to those in fig. 4 (a), visible differences are observed between the *k*-space maps generated using these two theoretical approaches.

Comparing the results of the one-step theory calculations to the experimental *k*-space maps we observe encouraging similarities. In particular, the sizes of the Brillouin zone features, and the general intensity variations across the image, with noticeable depression in intensities in the first Brillouin zone (including $e_g$ intensity loss in fig. 3(b),(c)) are well reproduced. Most importantly, the bulk-interface difference maps for the LSMO-derived Mn 3*d* $e_g$ and $t_{2g}$ states show a similar degree of suppression in intensities at the LSMO/STO interface – about 9.5% in theory compared to 4.5% in experiment. Thus, although we cannot yet claim fully quantitative agreement between experiment and theory, these results demonstrate a significant first step in the interpretation and use of SWARPES data.

In conclusion, by combining soft x-ray ARPES with standing-wave excited photoelectron spectroscopy, we have devised a unique technique for probing the $\vec{k}$-resolved electronic structure of buried layers and interfaces. By generating an x-ray standing wave inside a multilayer sample, and then translating it up and down within the sample by varying incidence angle, we can selectively probe electronic structure emphasizing the bulk of a layer and its interface, and then directly compare the two to each other. We have applied SWARPES to the investigation of the electronic properties of the buried interface within a magnetic tunnel junction $La_{0.7}Sr_{0.3}MnO_3/SrTiO_3$, and discovered that the bulk-like and interface-like regions of the buried $La_{0.7}Sr_{0.3}MnO_3$ layer exhibit distinctly different behavior, consistent with a change in the Mn bonding geometry at the $La_{0.7}Sr_{0.3}MnO_3/SrTiO_3$ interface observed previously [17], but now with $\vec{k}$ resolution. The experimental results are further validated via agreement with free-electron final-state model calculations and more precise state-of-the-art one-step photoemission theory including matrix element effects. Future theoretical treatments should involve the inclusion of atomic distortions near the interface, *e.g.* incorporating a crystal-field distortion near the interface that is suggested by our prior angle-integrated SW-XPS study of the same system [17] and more detailed calculations presented in our SI [34], as well as a more accurate inclusion of the standing wave intensity profile and phonon effects. We thus suggest that the SWARPES method should be of broad use in the future studies of buried layers and interfaces in various types of epitaxial multilayer structures, including those exhibiting spintronic, ferroelectric, multiferroic, and superconducting properties.

* * *

This research was supported by the U.S. Department of Energy, Office of Science, Office of Basic Energy Sciences, Division of Materials Sciences and Engineering under Contract No. DE-AC02-05CH11231, via both the LBNL Materials Sciences Division, Magnetic Materials Program, and the LBNL Advanced Light Source. J.M., J.B. and H.E are grateful for financial support from the German funding agencies DFG (FOR1346 and EB 154/20) and the German ministry BMBF (05K10WMA). D.D. and R.P. acknowledge funding by the German Science Foundation, SFB/TR80 and computational time at the Leibniz Rechenzentrum. Research at Stanford was supported through the Stanford Institute for Materials and Energy Science (SIMES) and the LCLS by the US Department of Energy, Office of Basic Energy Sciences. A portion of this work was performed using the TEAM 0.5 microscope at the National Center for Electron Microscopy (NCEM), which is supported by the Office of Science, Office of Basic Energy Sciences of the U.S. Department of Energy under Contract No. DE-AC02-05CH11231. The authors would also like to acknowledge the critical TEM sample preparation work performed by Marissa Libbee at NCEM.

**Figure Legends**

**Fig. 1.** Experimental setup and basic principles of standing-wave ARPES (SWARPES). (a) Schematic diagram of the ARPES experiment illustrating the angular degrees of freedom for the sample manipulation ($\theta_{TOA}$ and $\beta_{TOA}$), the hemispherical electrostatic photoelectron analyzer, the position-sensitive multichannel plate (MCP) detector with two orthogonal axes x = $k_x$ and y = $E_{kin}$, and the final CCD screen with the resulting $E_{kin}$ vs $k_x$ dispersion. The angle between photon incidence and the spectrometer lens axis was 60°, with both directions lying in the x-z plane. (b) Schematic diagram of the investigated multilayer structure consisting of 120 bilayers of STO and LSMO grown epitaxially on a single-crystal STO substrate, with each bilayer consisting of 4 units cells (15.61 Å) of STO and 4 unit cells (15.51 Å) of LSMO. A photon energy of 833.2 eV corresponding to the maximum reflectivity at the La $3d_{5/2}$ absorption edge was used for the photoemission experiments [17]. An example of the $E_{kin}(\vec{k})$ distribution for a fixed value of binding energy $E_B$ is shown above the sample. (c) Standing-wave excited photoemission intensity rocking curves for Ti $2p_{3/2}$ and Mn $3p$ core-levels (solid curves), as well as the x-ray optical simulations fitted to them (dashed curves), and yielding previously the chemical depth profile of the sample [17]. (d) Simulated intensity of the x-ray standing-wave electric field ($E^2$) inside the sample as a function of depth and grazing incidence angle. The line cuts indicate that, for incidence angles < 12.4°, the standing wave field highlights the bulk or center of the LSMO layer, but for angles >12.9° the interface regions of the LSMO layer are emphasized.

**Fig. 2.** Separating band-dispersions in SWARPES from residual x-ray photoelectron diffraction. (a) A typical ($k_x$, $k_y$) map for a fixed value of binding energy $E_B$, integrated over a 300 meV window containing the the Mn $3d$ $t_{2g}$ valence states, including an estimated 25% of intensity due to x-ray photoelectron diffraction (XPD). (b) A corresponding ($k_x$, $k_y$) XPD map of the Mn $3p$ core-level

exhibiting only XPD modulation. (c) A simulation of the Mn 3*p* XPD pattern using dynamical diffraction (Kikuchi-band) theory. (d) The corrected ($k_x$, $k_y$) map obtained by normalizing the combined Mn 3*d* $t_{2g}$ + XPD spectrum in (a) by the XPD spectrum in (b), so as to more clearly obtain the dispersive electronic structure of the Mn 3*d* $t_{2g}$ states, via method described in detail in our SI [34].

**Fig. 3.** Depth-resolved SWARPES measurements of the LSMO/STO superlattice. (a) An angle-integrated spectrum spanning the binding-energy window of 10 eV (from +1 eV to -9 eV), and including all the major features of the valence-bands, labeled 1 – 5, with their origins and characters indicated. (b) SWARPES data for these five energies in a bulk-LSMO sensitive SW measurement geometry. Binding-energy integration windows of 300 meV (consistent with our total energy resolution) centered around the binding energies of the five features discernible in the angle-integrated valence spectra (as determined by peak-fitting), were used to obtain these plots. Shown are XPD-normalized angle-resolved ($k_x$, $k_y$) photoemission intensity maps of the Mn 3*d* $e_g$ (1), Mn 3*d* $t_{2g}$ (2) states, the largely STO-derived states (3 and 4), and the valence-band bottom states (5) due largely to LSMO. **(**c) As (b), but for an LSMO/STO-interface sensitive measurement geometry of the SW. (d) Bulk – interface difference ($k_x$, $k_y$) maps based on (b) and (c), revealing the most significant differences for the LSMO-derived Mn 3*d* $e_g$ and Mn 3*d* $t_{2g}$ states at the interface between STO and LSMO, and as well as the dispersive valence-band bottom bands 5 from LSMO. The intensity scales at right indicate the relative amplitudes of the effects.

**Fig. 4.** Theoretical calculations for SWARPES from levels 1 - Mn 3*d* $e_g$, 2 - Mn 3*d* $t_{2g}$ and 5 – the bottom of the valence bands. **a,** Simple free-electron final-state theory with direct transitions from an LDA+U-based band structure (see our SI [34]). Yellow color corresponds to spin-up (majority) bands, and red to spin-down (minority). (b), (c), and (d) More accurate one-step photoemission theory summing over both spin polarizations and with the standing-wave intensity profile included, for (b)**,** a bulk-LSMO sensitive

geometry, (c), an interface-sensitive geometry, and (d), the bulk-minus-interface difference, respectively. The amplitudes of effects are again indicated. (e) represents a direct comparison to experimental panels from fig. 3 (d).

**FIG. 1**

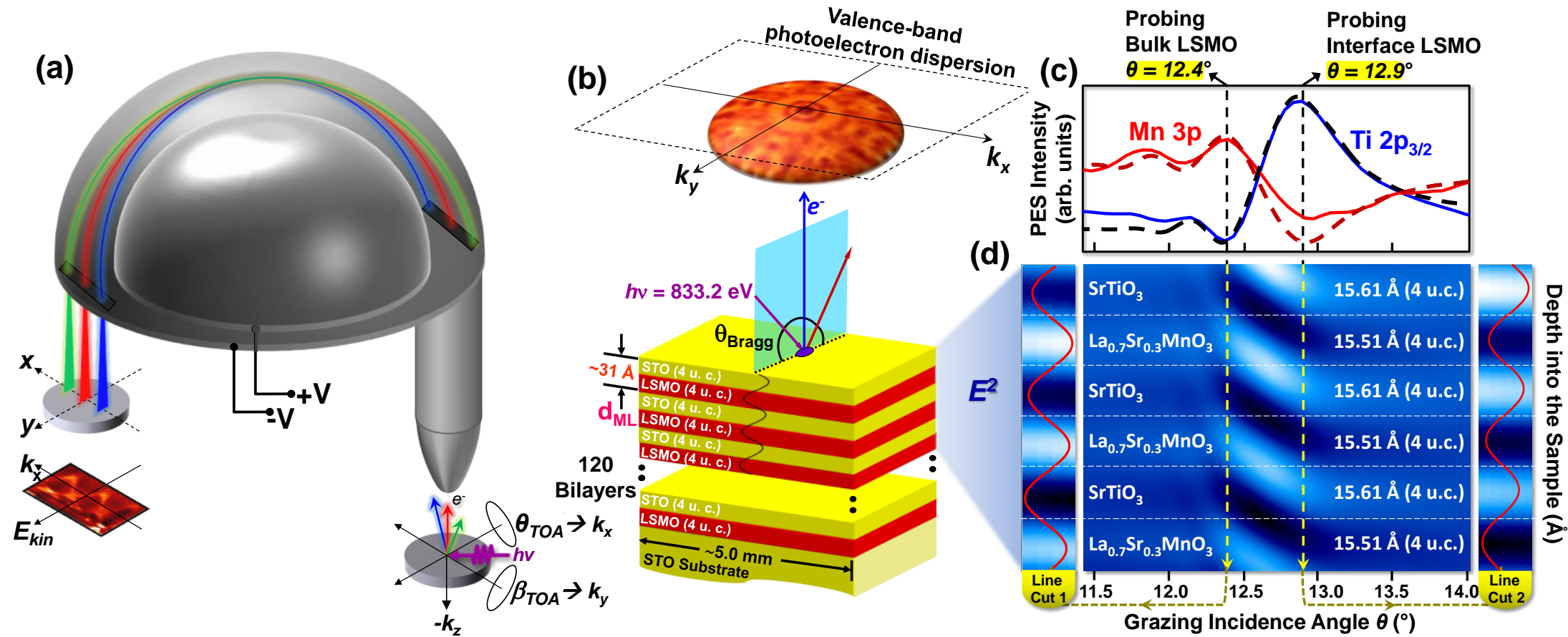

**FIG. 2**

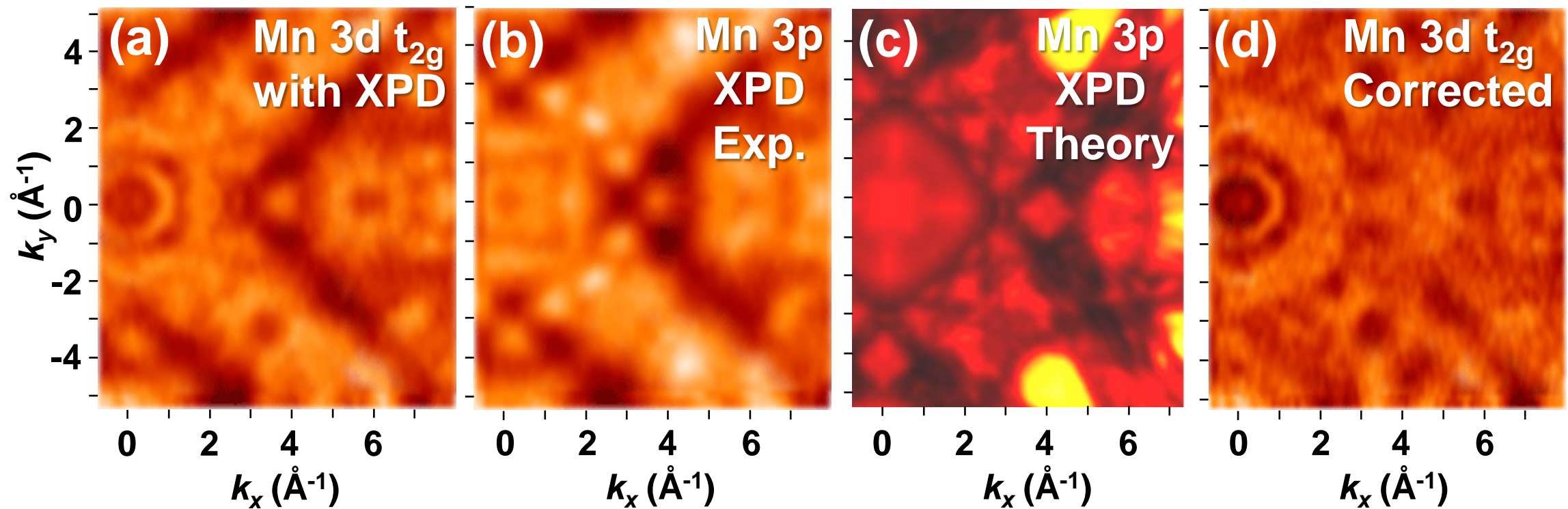

**FIG. 3**

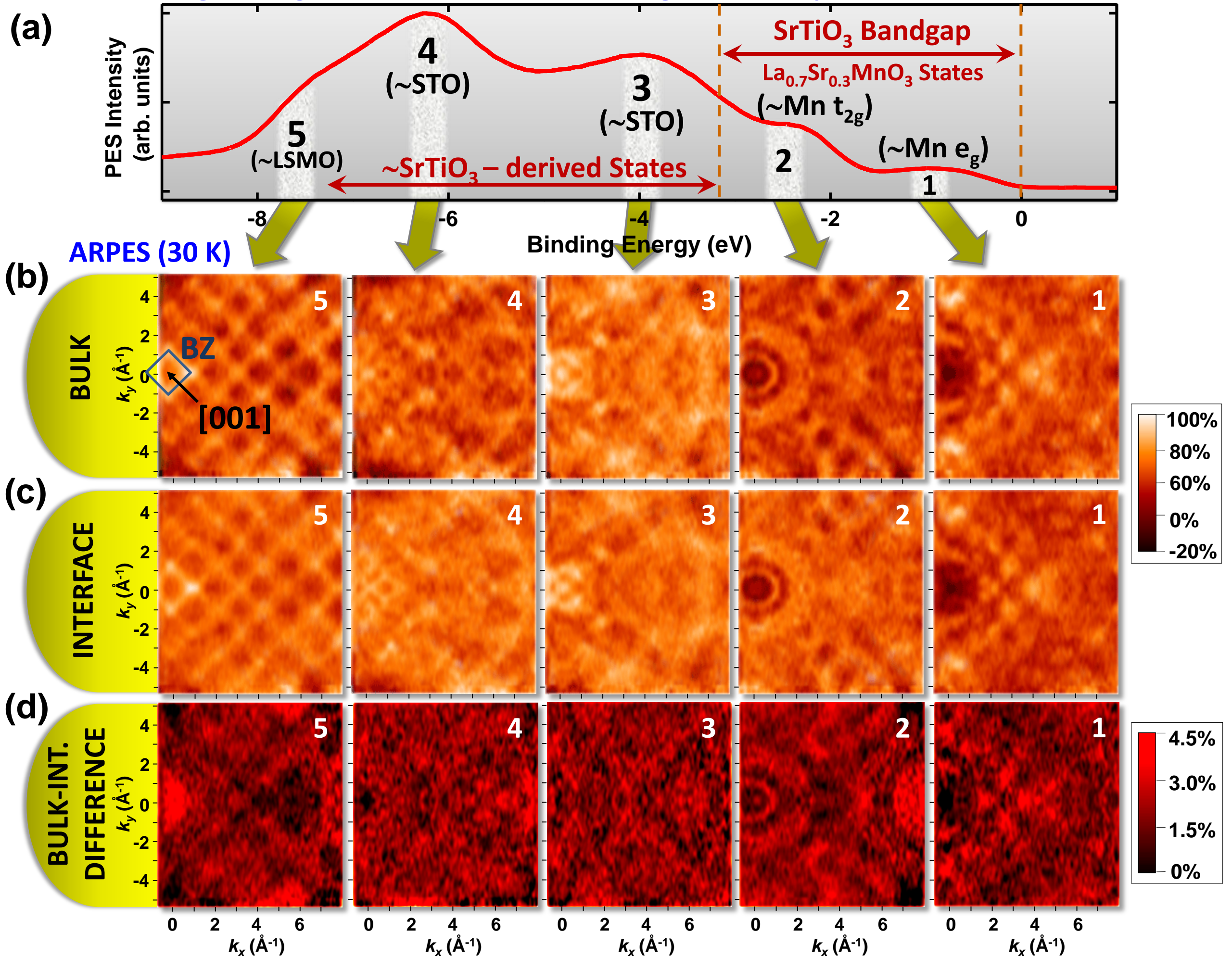

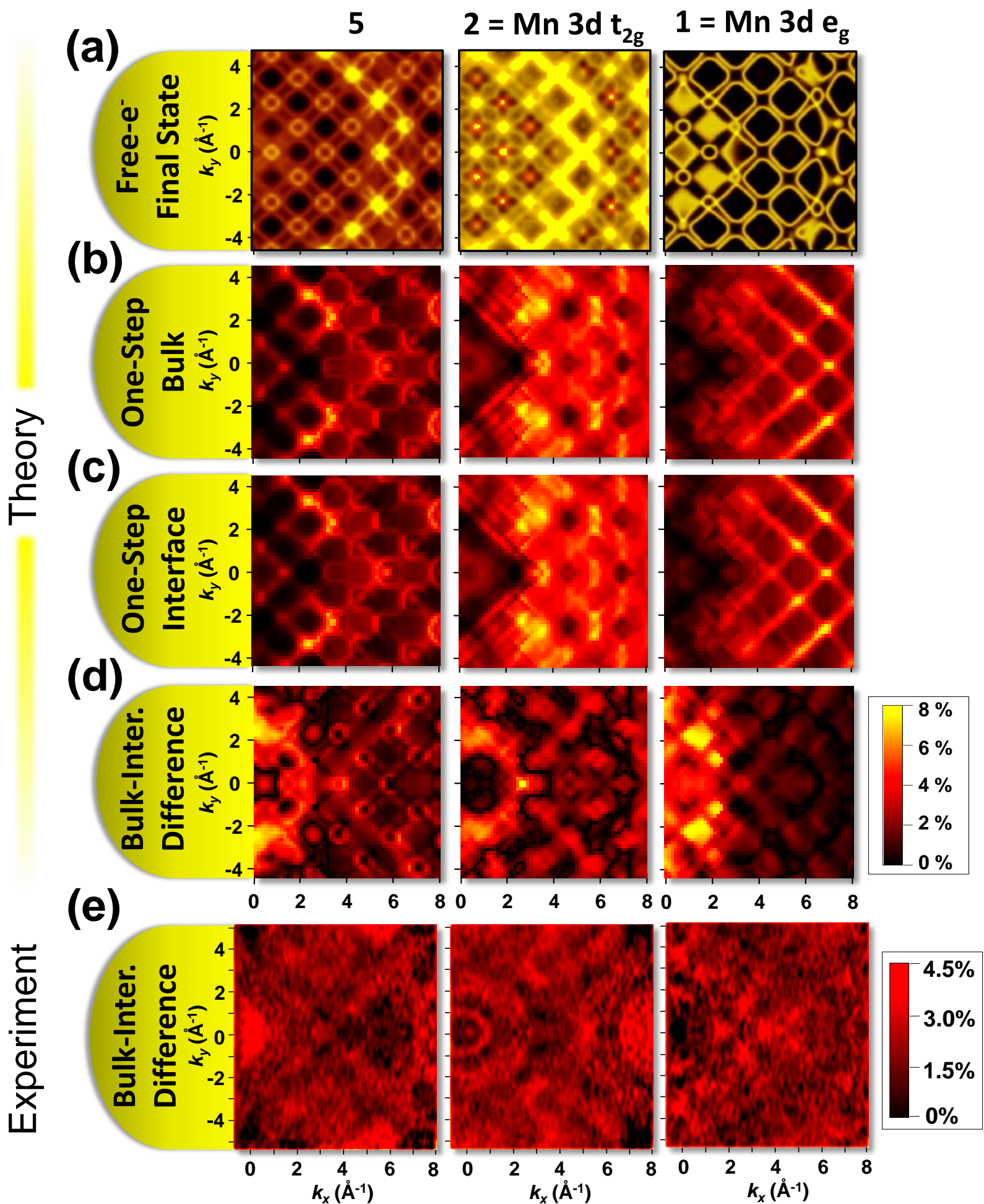

FIG. 4

# Momentum-resolved electronic structure at a buried interface from soft x-ray standing-wave angle-resolved photoemission

We here present some supplementary materials related to our method of sample growth, measurement technique, data analysis, the several levels of theoretical modeling we have done, and the more detailed properties or our sample. This includes our method of correcting for x-ray photoelectron diffraction modulations in the ARPES results, as well as some additional theoretical results to clarify the basic electronic structures of LSMO and STO and what we expect to see in experiment, some fully self-consistent LDA+U results for the multilayer, and further clarifications concerning the one-step photoemission calculations that are our major interpretive tool. These provide additional insights into the validity and interpretation of our experimental data. In addition, we present supplementary characterization measurements of our LSMO/STO superlattice sample, which include magnetization, electrical transport, and high-resolution cross-sectional STEM analysis with imaging by both high-angle annular dark field (HAADF) and electron energy loss spectroscopy (EELS).

**Superlattice growth and characterization**

The LSMO/STO superlattice sample consisted of 120 bilayers, nominally consisting of 4 unit cells of LSMO (~15.51 Å) and 4 unit cells of STO (~15.61 Å), and was fabricated using the PLD technique, with reflection high-energy electron diffraction (RHEED) for monitoring the growth process. Atomically smooth $TiO_2$-terminated STO(100) substrates were prepared by a combined HF-etching/anneal treatment [1]. All substrates had vicinal angles of ~0.1°. A stoichiometric LSMO target and a single-crystal STO target were ablated at a laser fluence of 1.5 J/cm$^2$ and a repetition rate of 1 Hz. During growth, the substrate was held at 750 °C in an oxygen environment at $2.6 \times 10^{-1}$ mbar. The growth process was optimized in a previous study so as to result in an ideal unit-cell-controlled layer-by-layer growth and bulk-like magnetic and transport properties [2]. A low level of surface roughness (maximum of 4 Å) was confirmed by atomic force microscopy (AFM). Out-of-plane superlattice periodicity was confirmed to be 31.13 Å using x-ray diffraction (XRD), very close to the expected bilayer thickness of 31.12 Å. An analysis of rocking curves in a prior SW-XPS study of a similar multilayer confirmed the high quality, although noting a decrease in the bilayer period as growth continued [3], an effect confirmed below by TEM. Clear ferromagnetic behavior was observed up to room temperature in a SQUID Magnetometer.

**Angle-resolved photoemission measurements**

The photoelectrons were analyzed by means of a hemispherical analyzer (VG Scienta R4000) equipped with a two-dimensional microchannel plate (MCP) detector [4]. A six-axis sample manipulator permitted rotations in both the take-off angle $\theta_{TOA}$ and the orthogonal angle $\beta_{TOA}$. Small corrections to the incidence angle $\theta_{inc}$ due to the rotation in $\beta_{TOA}$ were also made. The Fermi level was frequently calibrated using a Au reference sample. In presenting detector images, ~2.5° of the detector angle range (corresponding to ~0.75 Å$^{-1}$ in $k_x$) on both sides of each ($k_x, k_y$) map was cropped in order to remove experimental artifacts associated with the detector edges.

It might be noted that the angular positions of the Bragg features for the rocking curves in Fig. 1(c) deviate somewhat (by ~1°) from those reported in our previous study of a similar, but thinner, multilayer sample (48 LSMO/STO bilayers, compared to the 120 bilayers of the sample in the current study) [ref. 17 in the paper]. This deviation is due to a combination of the variation in the bilayer thickness with successive layers, as discussed in a prior study [3] below in connection with the TEM results, and an improved procedure for calibrating the incidence x-ray angle. The angles here are thus more accurate.

**Correction of data for x-ray photoelectron diffraction and density-of-state effects, and bulk-interface difference maps**

In correcting our SWARPES raw data for the effects of x-ray photoelectron diffraction and density-of-states effect induced by phonon-induced non-direct transitions, each detector image in ($k_x, k_y$) was first divided by the average over angle and average over energy, as described in the text [5,6]. Then, to remove any remaining XPD contributions, which are estimated to be about 25% of intensity based on a Debye-Waller factor calculation, we performed a scaled normalization of the combined Mn $3d$ $t_{2g}$ + XPD $(k_x, k_y)$ maps using angle-resolved spectra of the Mn $3p$ core-level at a binding energy of 42.2 eV and a kinetic energy very close to the valence photoelectrons (~791 eV compared to ~833 eV): (1) Averages of the valence-band $\vec{k}$-space maps for more localized flat-band and XPD-like bands 3 and 4 at the binding energies of 4.0 eV and 6.2 eV were taken, and several brightest and darkest diffraction spots on these averaged VB maps were selected (4 brightest and 3 darkest). (2) Mn $3p$ XPD maps were scaled such that the same diffraction spots on these maps match the intensities of the selected diffraction spots on the averaged valence-band maps for each experimental geometry. (3) These scaled Mn $3p$ maps were then used to normalize the valence-band maps for each corresponding experimental geometry by simple subtraction, thus assuming a linear addition of bandlike dispersive effects and XPD effects. Such a normalization procedure also removes any signature of the 2D detector non-uniformity, a purely instrumental effect.

Additional aspects of the analysis of our data make it clear that this correction was reliable and did not introduce any artifacts in the final LSMO bulk-minus-interface SWARPES difference results. In fig. S1, we first show the complete set of angles at which SWARPES data was obtained, indicated on top of the core-level rocking curves that were used to determine the most bulk LSMO and most LSMO/STO interface sensitive angles, denoted as C and E on the figure. From the standing wave plot in fig. S1(c), we can thus say that angles A, B, C should be more interface sensitive and angles E, F, G more bulk sensitive, with D being somewhere in between.

In figs. S2-S4, we show the bulk-interface differences of SWARPES data for different pairs of angles, so as to either emphasize the difference, or minimize it by looking at two bulk or two interface angles. All of the data has here been corrected for XPD effects using the procedure described above. Fig. S2 represents two angles expected to be sensitive to bulk LSMO, and these differences show essentially no discernible fine structure. Fig. S3 is for the two angles showing maximum standing-wave contrast between bulk and interface, as shown already in fig. 3(d) of the main text; clear differences are seen for the $e_g$, $t_{2g}$, and valence-band bottom panels that we finally assign to LSMO, which are expected to show the biggest effects. Fig. S4 represents two angles expected to be sensitive to the LSMO/STO interface, and here again, there is no discernible fine structure. We thus conclude that we are able to reliably measure the ca. 5% effects that are differences between the bulk and the interface electronic structure.

**Reference band structure calculations and first-order simulations:**

Band structure calculations for the band insulator STO and the half-metallic ferromagnet LSMO, as calculated with the Wien2k program in the LDA (for STO) and LDA+U (for LSMO) approximation [7], are shown in fig. S5, (a) and (b). For LSMO, we have used an effective U equal to the Mn 3*d* onsite Coulomb parameter U minus the Mn exchange parameter J of 2.0 eV. In fig. S5(c), the orbital-projected densities of states for LSMO are also shown. The STO conduction-band states have been shifted using the so-called scissor operator so as to yield the experimental 3.3 eV indirect band gap. The two band structures have also been shifted relative to one another by the experimental valence-band offset of 3.0 eV, as measured using a standard x-ray photoemission (XPS) technique based on valence-band and core-level spectra excited by hard x-rays from our multilayer and from the two bulk materials STO and LSMO of which it is made [8,9]. This band offset is also in good agreement with a 2.73 eV value calculated within LDA+U for the actual 4 unit cell/4 unit cell multilayer structure, using the all-electron method described in the next section. This plot makes it clear that we can expect to see Mn $e_g$ and $t_{2g}$ derived states over ca. 0-3 eV binding energy. At this level of bulk theory, it is not clear whether the valence band minimum of STO or LSMO will be lower in energy, but we clarify this below.

The expected contributions of LSMO and STO versus binding energy in our spectra have also been estimated through these bulk densities of states by assuming that each LSMO or STO layer of thickness *t* in the multilayer contributes an intensity proportional to *[1-exp(-t/Λ$_e$sinθ$_{TOA}$)]*, where *Λ$_e$* is the inelastic mean free path in that layer (as estimated from the TPP-2M formula [10]), and that this intensity is reduced by all overlying layers of thickness t' according to *exp(-t'/Λ$_e$sinθ$_{TOA}$)*. Each orbital-projected layer DOS has also been multiplied by atomic differential photoelectric cross sections to approximately allow for matrix-element effects. The results of these calculations are shown in fig. S6, where they are compared to angle-integrated experimental spectra that should approximate matrix-element weighted densities of states; these are shown for the two angles that maximize sensitivity to the LSMO/STO interface and bulk LSMO. These calculations further confirm that we expect to see LSMO bands over ca. 0-3 eV, and that STO will dominate at ~75% or more for larger binding energies. The relative intensities of peaks 1 and 2 in the experimental data also nicely confirm the enhancement of the LSMO-derived features when the standing-wave moves to the central LSMO bulk-sensitive position, further verifying our overall methodology.

**Fully self-consistent electronic structure calculations for the multilayer:**

For a more accurate look at the electronic structure of our sample, we have also carried out fully self-consistent density functional theory calculations on LSMO/STO superlattices of our ideal 4 unit cell/4 unit cell configuration (see fig. S7(a) using the all electron full-potential augmented plane-wave method in the WIEN2k implementation [7]. Electronic correlations beyond the generalized gradient approximation (GGA) [11] were considered in the LDA/GGA+U method [12] with $U = 3.0$ eV, $J = 0.7$ eV on Mn $3d$ states, $U = 5.0$ eV, $J = 1.0$ eV on Ti $3d$ states and $U = 8.0$ eV, $J = 0.0$ eV on La $4f$ states. The statistical distribution of La and Sr was treated in the virtual crystal approximation. The rhombohedral LSMO bulk structure was also transformed to monoclinic to fit on the $SrTiO_3(001)$ substrate and the lateral lattice parameter of the superlattice was set to the GGA-lattice constant of $SrTiO_3$ ($3.92 \times \sqrt{2}$ Å). Full relaxation of the internal structural parameters was performed in an 80-atom unit cell, allowing for all octahedral tilts and rotations.

Some of these results are shown in fig. S7, including in (b) the spin-resolved layer-by-layer total density of states, with Mn-containing layers representing the two interfaces (IF) and the two internal (IF-1) or bulk LSMO layers indicated. The energy locations 1-5 of the points at which we have chosen to present SWARPES results are also indicated. These calculations indicate that the changes in DOS from the interface to internal/bulk regions of LSMO are subtle, but certainly present. From the atomic identities in the idealized layer structure shown, it is also clear that the top and bottom interfaces are not identical, and the consequences of this are also evident in the TEM results we show below. The projected Mn densities of states shown in fig. S7(c) also indicate marked differences between the multilayer and bulk LSMO, as well as between interface and internal/bulk LSMO layers in the multilayer. These results also indicate that the intensity from the SWARPES results at energy 5, the bottom of the valence bands, should arise predominantly from LSMO, as we have modeled it in our free-electron final-state calculations (cf. fig. 4(a) in the main text). The difference in this aspect from the theoretical results in fig. S5 is no doubt due to the more complete set of U and J parameters used here, thus better allowing for correlation effects. The bandgap of STO is also better predicted in fig. S7 for the same reason. Optimization of atomic positions also indicates in results not shown here that there is a Jahn-Teller effect in the interface octahedra that is fully consistent with a prior core-level SW-XPS study [3], and that the octahedra are compressed along z, rather than elongated, and show a strong orbital polarization of ~8.4%, with stronger occupation of $d_{x2y2}$; this polarization is also found to be much smaller (~1.7%) and of opposite sign in the internal/bulk layers. Future SWARPES measurements with variable light polarization should permit directly measuring such effects.

**One-step theory of photoemission calculations:**

Self-consistent electronic structure calculations were first performed within the ab-initio framework of spin-density functional theory. The electronic structure was calculated in a fully relativistic mode by solving the corresponding Dirac equation. This was achieved using the spin polarized relativistic multiple-scattering or Korringa-Kohn-Rostoker formalism [13]. To account for electronic correlations beyond the LSDA [14] we employed a LSDA+U scheme as implemented within the relativistic SPR-KKR formalism, including for LSMO the average screened Coulomb interaction *U* (an adjustable parameter, chosen as $U_{Mn}$ = 2.0 eV) and the Hund exchange interaction *J* (calculated directly and set to $J_{Mn}$ = 0.9 eV) [15] and for STO $J_{Ti}$ = 0.9 eV. Substitutional disorder has been treated within the coherent potential approximation, which is considered to be the best available single-site alloy theory. The effective potentials were treated within the atomic sphere approximation (ASA). A sample consisting of repeated 4 unit cells of LSMO/4 unit cells of STO was used to calculate self-consistently the electronic structure of the corresponding semi-infinite half-space, thus yielding the effective potentials. Our photoemission calculations are based on these electronic structure inputs. Lifetime effects in the initial and final states have been included via imaginary values of the potential $V_{i,f}$. To take care of impurity scattering a small constant imaginary value of $V_{ii} = 0.08$ eV was used for the initial states. For the final states, a constant imaginary part $V_{if} = \sim 3.0$ eV has been chosen to simulate the IMFP for our photon energy, corresponding to an inelastic mean free path for intensity of about 19 Å. Furthermore, the layer-resolved photocurrent was weighted layer-by-layer with the corresponding electric-field intensity $|E^2|$ profile of the standing wave inside the superlattice as a function of depth and incidence angle as derived from our optical model (cf. fig. 1(c) in the main text). Finally, the current was averaged over a 300 meV energy window, which corresponds to the experimental data binning.

We note here that the band structures and densities of states initially calculated in deriving the atomic potentials for the SPR-KKR method agree well with those in fig. S6. A further important point is that the full multilayer was included in these calculations, so that the effects of scattering of electrons originating in the LSMO layers by the STO interlayers and final STO overlayer were explicitly included. However, since phonon effects leading to non-direct transitions and DOS+XPD effects [16,17] were not included in the calculations, it is still appropriate to have corrected our experimental data for these effects. Also, no allowance was made for the relaxation of atomic positions near the interface, although this is an obvious point for future investigation and we have in the separate set of calculations described above begun to explore this.

**Magnetization and Electrical Transport Measurements:**

Electrical transport properties of the sample were measured using the four-point-probe technique, and are shown in fig. S8(a). The magnetic properties of the sample were measured in a Quantum Design

SQUID Magnetometer (MPMS). Figs. S8(b),(c) show the temperature dependence of the saturation magnetization (b) and typical magnetization curves at 10 K and 290 K (c) along the [100] direction after magnetic field cooling at 1 Tesla from 360 K. The values of $T_c$, resistivity and saturation magnetization are all consistent with prior studies of LSMO/STO multilayers in this thickness range [18].

**High-Resolution Cross-Section STEM Measurements:**

In order to verify with a direct imaging technique our previous standing-wave rocking-curve analysis of concentration depth profiles in a similar STO/LSMO multilayer, as reported previously in ref. 17 of our letter [3], we have performed high-resolution cross-sectional STEM measurements with both high-angle annular dark field (HAADF) and electron energy loss spectroscopy (EELS), using the aberration-corrected TEAM 0.5 microscope at the National Center for Electron Microscopy. utilizing a remote operation computer station (RemoTEAM) located at the Electron Microscopy Center at Argonne National Laboratory. The results of these measurements are shown in fig. S9 below. From the quantitative analysis of intensities of over 800,000 atomic columns across the full 120 bilayer cross-section, we have measured the interfacial roughness/interdiffusion between the LSMO and STO layers to be 1 - 1.5 u.c. This result is further confirmed by directly measuring the EELS chemical signal ratio of Ti/Mn in several smaller regions of the film, with the LSMO-on-STO interface exhibiting slightly more interdiffusion/roughness than the STO-on-LSMO interface. Furthermore, the layer thickness is directly measured to be 4.0 unit cells at the beginning of the growth process, but is found to decrease to 3.4 unit cells at the 120$^{th}$ bilayer. Thus, the STEM results fully confirm our prior SW-XPS finding that there is a gradient in the thickness of the STO and LSMO layers from top to bottom of the superlattice [3], and attest to the accuracy of the multilayer optical constants that we have used for our simulations.

# Figure S1

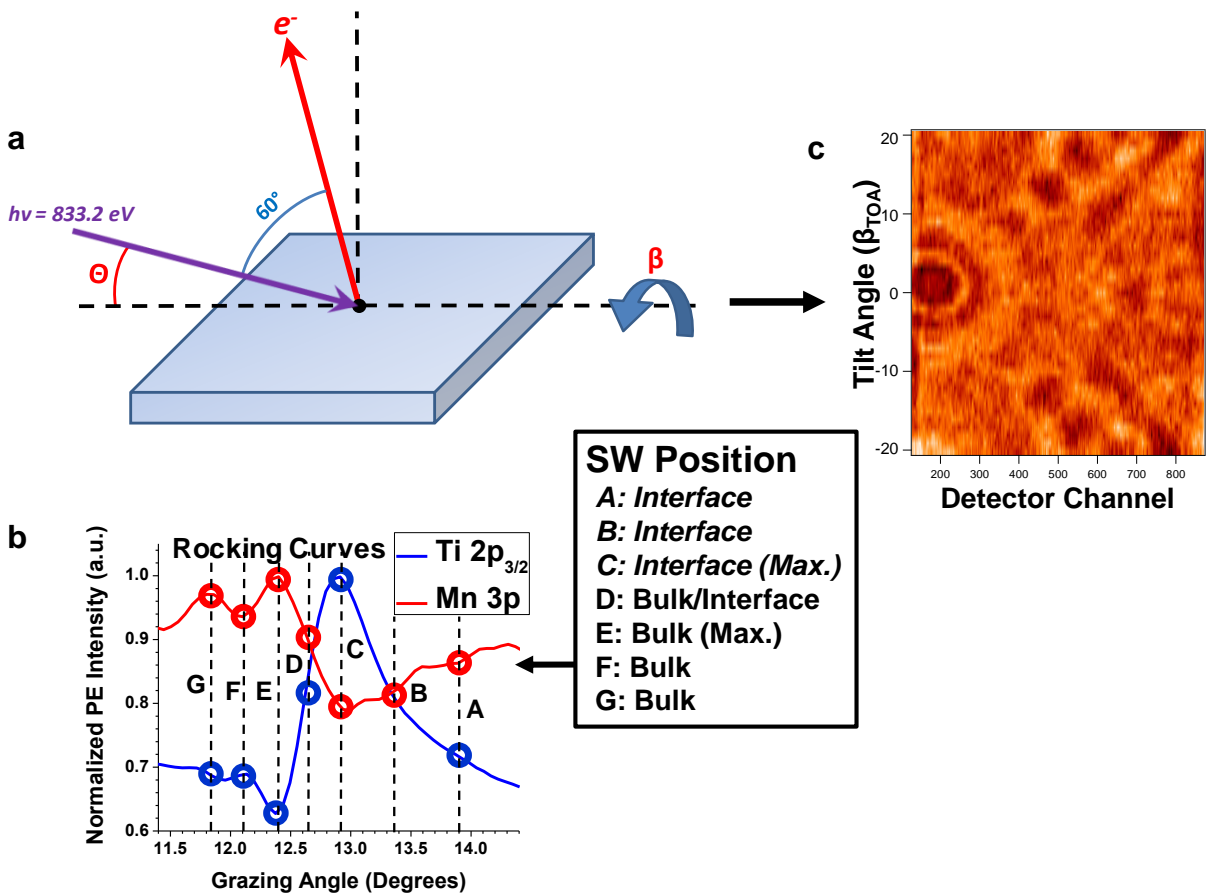

**Fig. S1.** (a) Our measurement geometry. (b) The seven different grazing incidence angles (Θ) at which SWARPES data were collected, corresponding to various positions along the Ti 2$p_{3/2}$ and Mn 3$p$ rocking curves. (c) For each value of Θ, an ARPES measurement was collected at 40 different tilt angles (β$_{TOA}$) ranging from -20° to +20°, finally yielding corrected data as shown here for the example energy at point 2 expected to be dominated by Mn $t_{2g}$ states.

# Figure S2

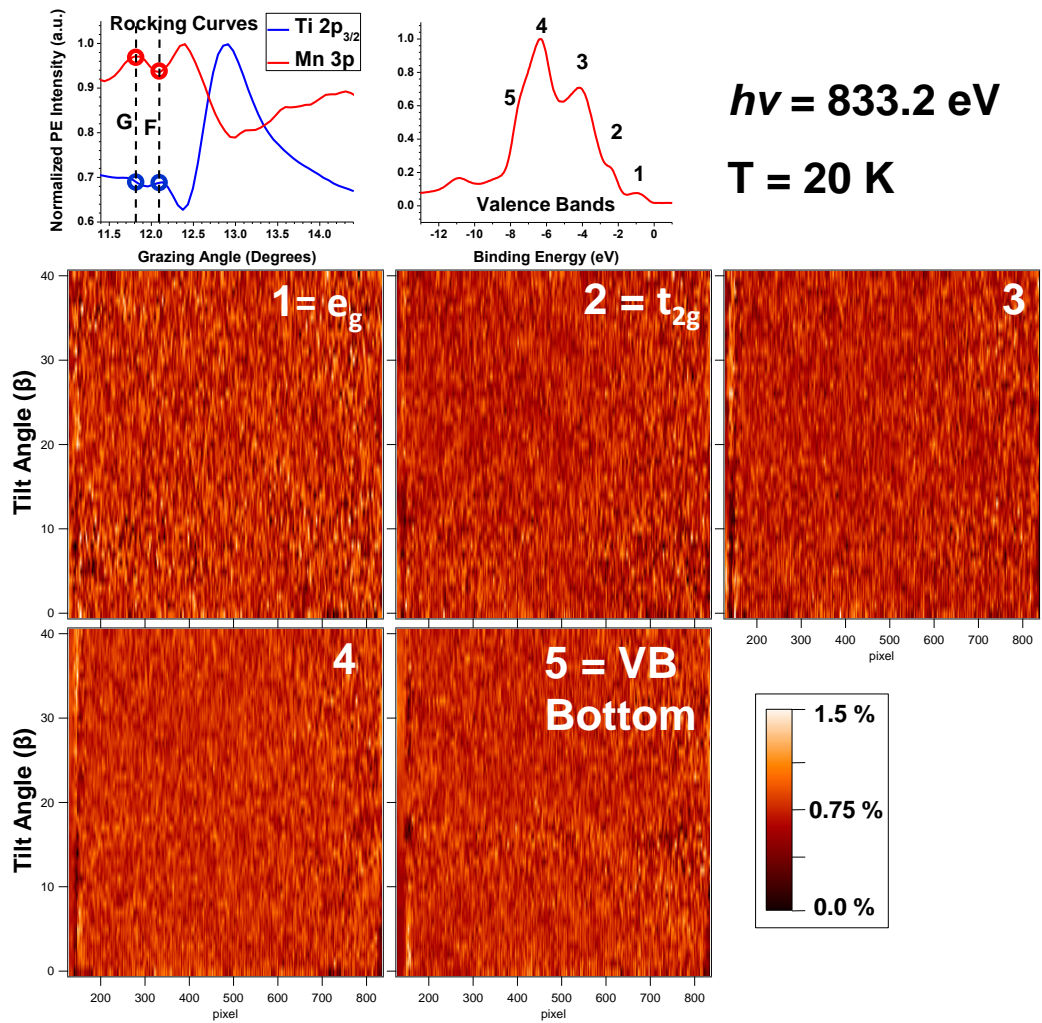

**Fig. S2.** Difference SWARPES patterns between angles F and G that should both be more sensitive to bulk or central LSMO, with the angular positions relative to the Ti $2p_{3/2}$ and Mn $3p$ rocking curves and energy positions in a density-of-states shown in the top two panels. The relative magnitude of the intensity modulations is also indicated.

# Figure S3

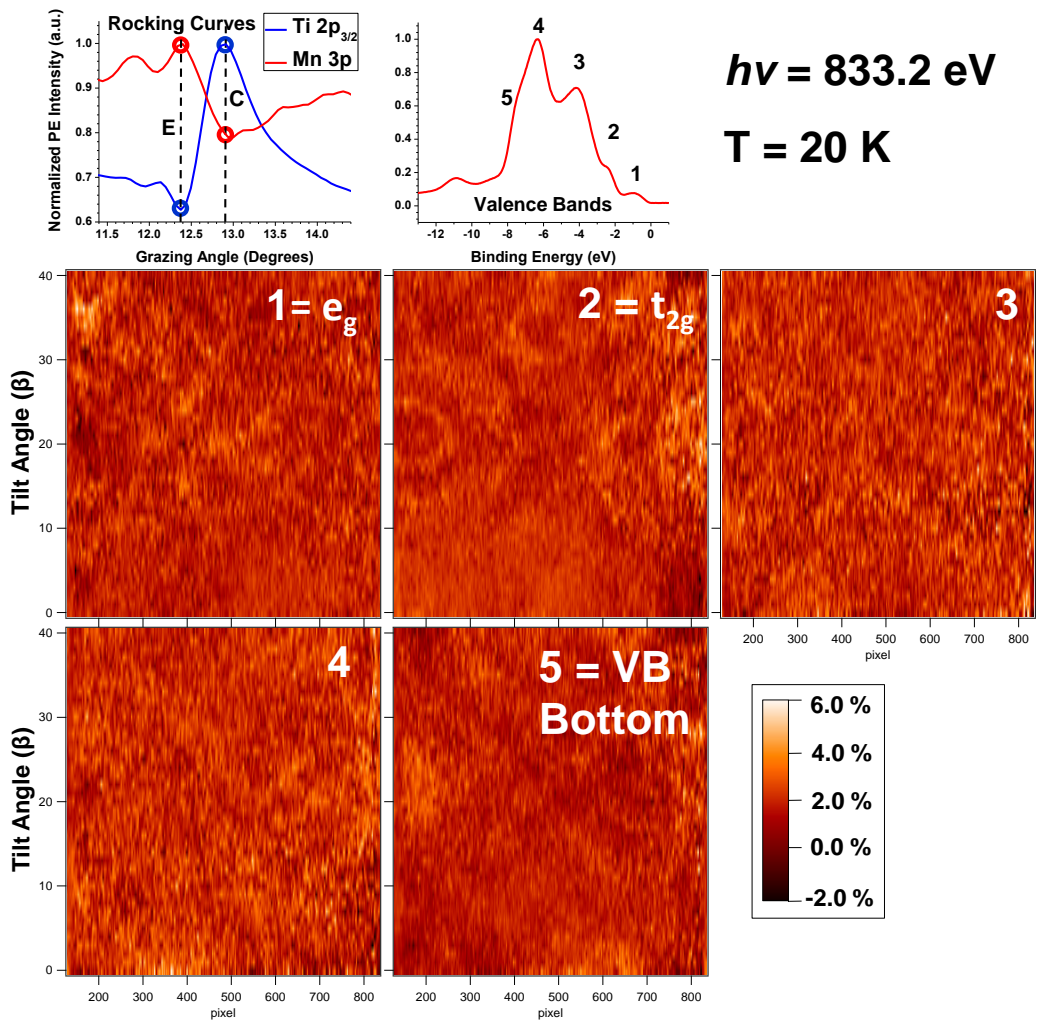

**Fig. S3.** As fig. S2, but for the angles C and E exhibiting maximum standing-wave contrast to bulk and interface.

# Figure S4

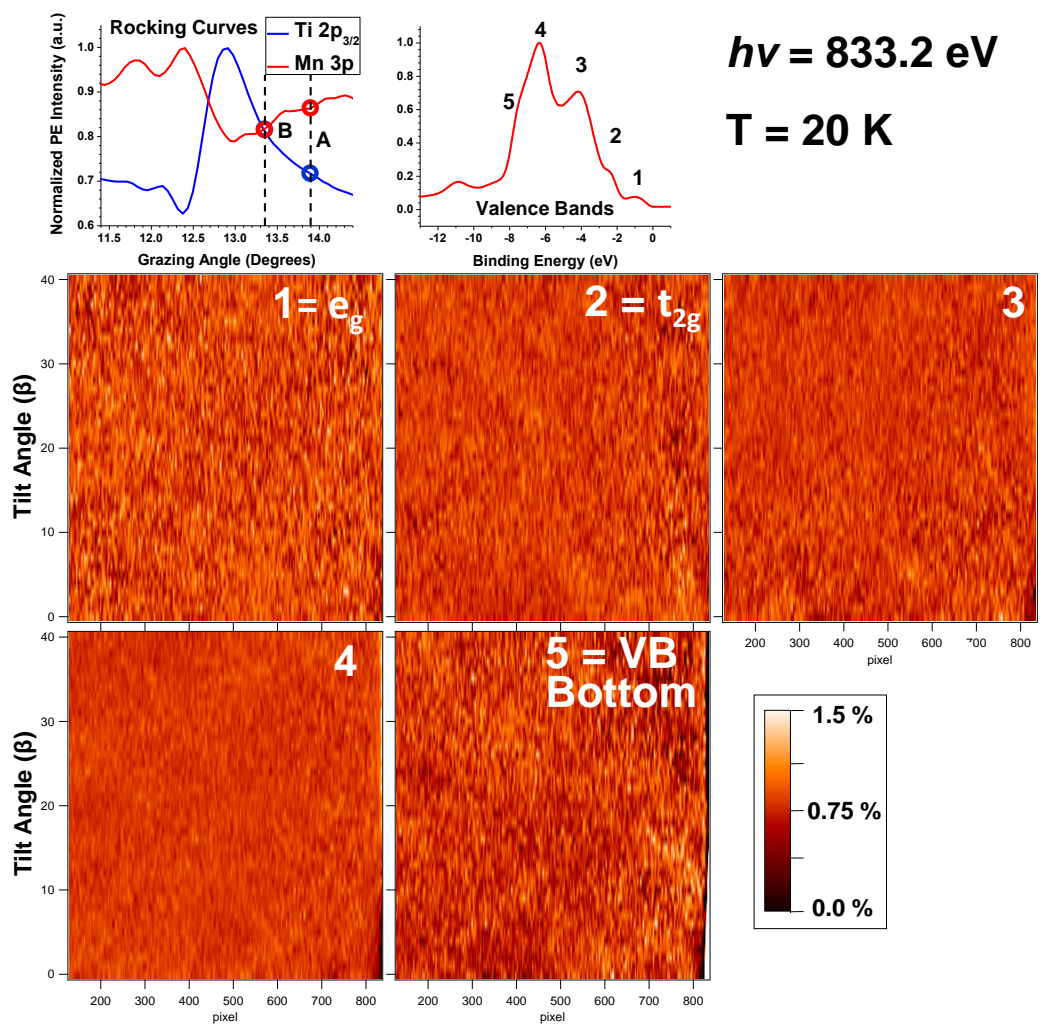

**Fig. S4.** As fig. S2, but for the angles A and B that should both be more sensitive to the LSMO/STO interface.

# Figure S5

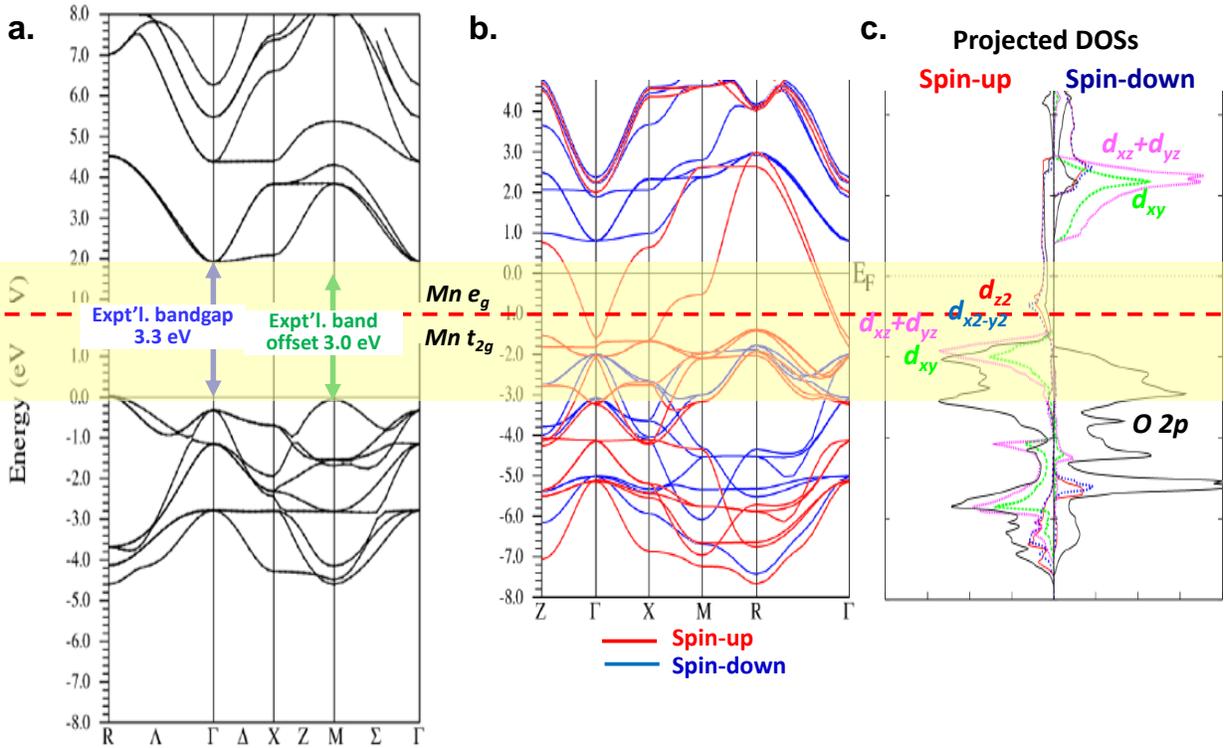

**Fig. S5.** The LDA band structures of (a) the band-insulator STO and (b) the half-metallic ferromagnet LSMO, as calculated with the Wien2k program. The band gap for STO has been adjusted with a scissors operator to agree with the experimental indirect bandgap. The experimental band offset in our sample has been measured using hard x-ray photoemission from core levels and valence bands [9]. The shaded yellow region is that over which the LSMO bands are expected to be seen in our SWARPES data. The calculations for LSMO were done in the LDA+U approximation. **c.** Projected densities of states for LSMO, indicating the expected $e_g$, $t_{2g}$ and O $2p$ makeup. LSMO is assumed to be ferromagnetic here, even though in our multilayer, it has only weak ferromagnetic order (cf. fig. S8).

# Figure S6

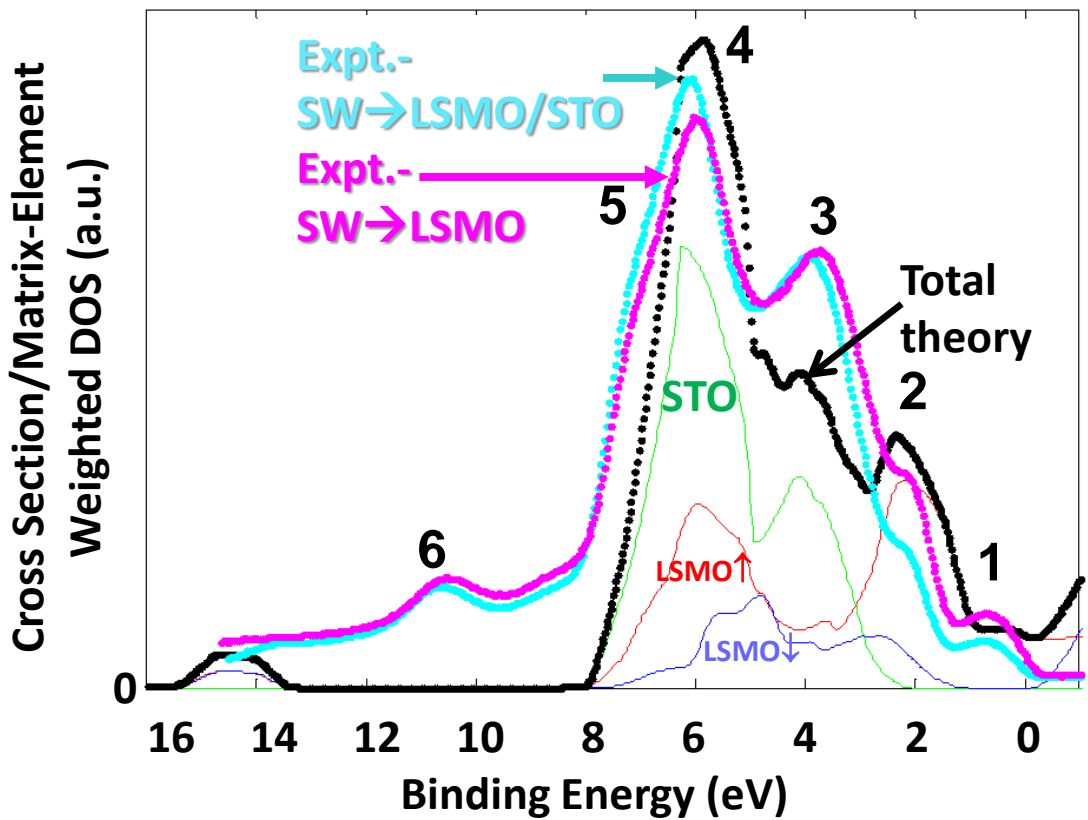

**Fig. S6.** Comparison of experimental angle-averaged DOS-like spectra at angles C (interface) and E (bulk) with LDA densities of states from the calculations of fig. S5 that have been summed with cross section and inelastic attenuation corrections over the 4 unit cell/4 unit cell structure of our multilayer. These show the dominance of LSMO over ca. 0-3 eV, and of STO for greater binding energies. Theory here is in error by about 4.5 eV for the O-2*s* dominated band 6 at about 10.5 eV in experiment, but this is not relevant to our SWARPES data.

# Figure S7

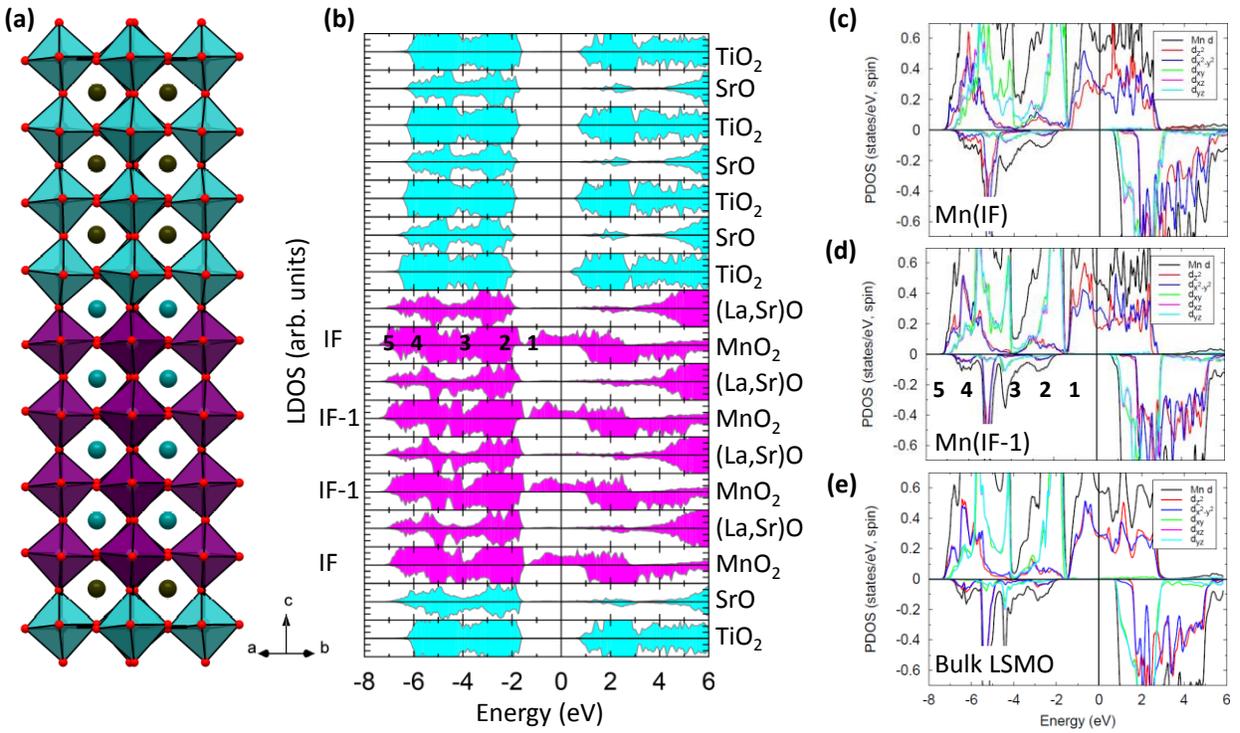

**Fig. S7.** Results from fully self-consistent all-electron GGA+U calculations for a 4 u.c. LSMO/4 u.c. STO multilayer. (a) Side view of the relaxed structure of the superlattice. (b) The layer-resolved total densities of states (LDOS), with Mn-containing layers at the interface (IF) and center "bulk" layers (IF-1) indicated. (c),(d) Projected densities of states for Mn summed over (c) the two interfaces IF, (d) the two center/bulk layers IF-1, and (e) for bulk LSMO. The energy positions at which we have chosen to show SWARPES results are indicated as 1-5 in (b) and (d). Note that the DOS due to the lowest-lying LSMO bands near the point R (see fig. S5) are expected to be below those of STO.

# Figure S8

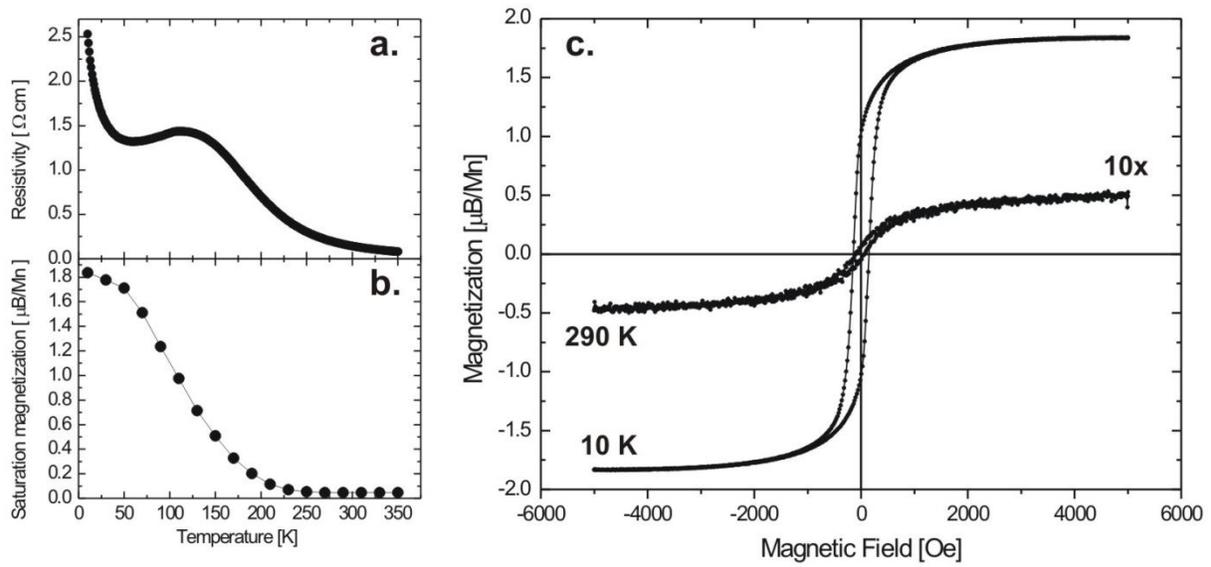

**Fig. S8.** Electrical transport and magnetic properties of the LSMO/STO superlattice. (a),(b) Temperature dependence of the resistivity and saturation magnetization respectively. (c) Magnetic hysteresis loops at 10 and 290 K showing ferromagnetic behavior.

# Figure S9

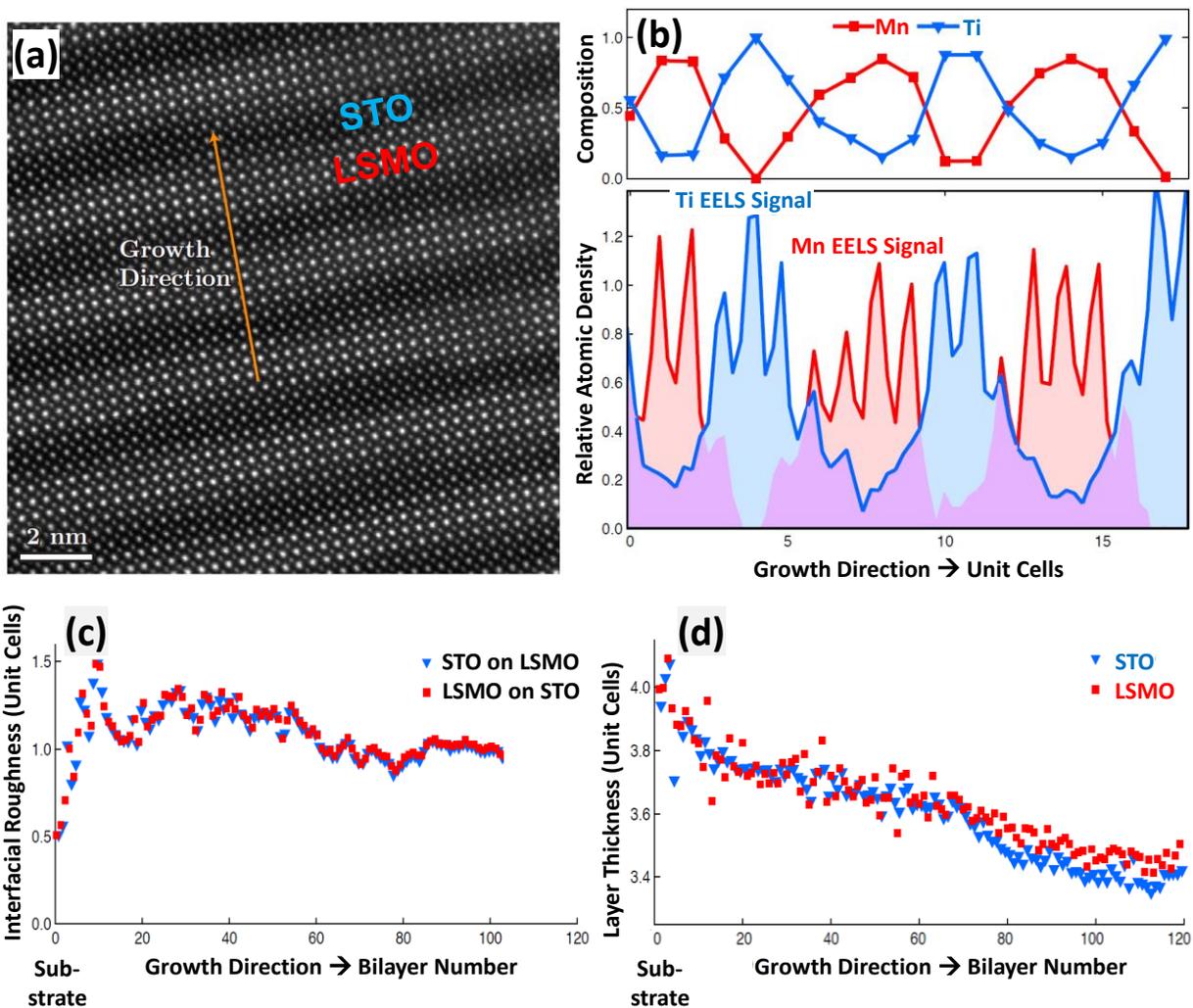

**Fig. S9.** (a) High resolution HAADF-STEM micrograph of the LSMO/STO sample in cross-section near the 90th grown layer, with the location of the EELS line scan indicated. (b) Ti/Mn column composition calculated from the integration of background-subtracted *L*-edge EELS. (c) Interfacial roughness calculated from the quantitative HAADF-STEM intensities of 800,000 atomic peaks across the full 120 bilayer cross-sectional sample. Roughness is defined as the RMS variance of the intensity midpoint of each bilayer transition, with each layer sampled over at least 200 nm. The steep drop-off after 100 layers is due to a slight bend in the TEM sample from residual stress in the multilayers, and should be ignored. (d) Layer thickness determined from the quantitative HAADF-STEM intensities of 800,000 atomic peaks across the full 120 bilayer cross-sectional sample. Layer thickness is defined as the peak-peak distance between maxima of an envelope function fit to atomic column intensity maxima.